\documentclass[letterpaper,twocolumn,american,prl,aps,superscriptaddress]{revtex4}
\usepackage[latin1]{inputenc}
\usepackage{amssymb}
\usepackage{amsbsy}
\usepackage{amsmath}
\usepackage{graphicx}
\usepackage{hyperref}
\usepackage[dvipsnames]{xcolor}

\def\bra#1{\langle #1|}
\def\ket#1{|#1\rangle}

\begin{document}
\title{Comment on: ``Fluctuation Theorem for Many-Body Pure Quantum States'' and Reply to ``arXiv:1712.05172''}

\author{Jochen Gemmer}
\affiliation{Department of Physics, University of Osnabr\"uck, D-49069 Osnabr\"uck, Germany}

\author{Lars Knipschild}
\affiliation{Department of Physics, University of Osnabr\"uck, D-49069 Osnabr\"uck, Germany}

\author{Robin Steinigeweg}
\affiliation{Department of Physics, University of Osnabr\"uck, D-49069 Osnabr\"uck, Germany}



\maketitle

 {\bf Comment on: ``Fluctuation Theorem for Many-Body Pure Quantum States''}
 
 Iyoda {\it et al.}\ present a general argument which is intended to (literal
 quotation of the first sentence of the abstract) ``\ldots prove the second
 law of thermodynamics and the non-equilibrium fluctuation theorem for pure
 quantum states'' \cite{iyoda17}. To exemplarily back up this statement, they
 perform a numerical analysis of a lattice model of interacting hard-core
 bosons.
 
 In this comment we point out the following: While the argument is mathematically
 sound, it can hardly be applied to the physical situations to which fluctuation
 theorems and/or the second law routinely refer. Its validity is limited to
 rather extreme time and length scales far away from said physical situations.
 This is due to Lieb-Robinson speeds being too high, relevant processes lasting
 too long, and baths being too small in standard setups. A careful analysis of
 the above numerical example reveals that the results supporting the theorem
 only occur for a selective choice of the parameters and/or coincide with the
 short initial period during which the system only negligibly leaves its initial
 state, i.e., the system practically ``does nothing''.
 
 Central to the argument in \cite{iyoda17} is a conceptual division of the bath
 $B$ into a near bath $B_1$ and a far bath $B_2$, in spite of the full physical
 bath just being one homogenous system. The considered system is only coupled
 to an edge of the near bath $B_1$. Now the claims in \cite{iyoda17} hold under
 two conditions: (i) The reduced state of $B_1$ must result as a strictly
 Gibbsian state for an energy eigenstate on the full bath $B$. (This corresponds
 to the eigenstate thermalization hypothesis (ETH).) (ii) All considered times
 must not exceed the minimum time at which energy, information, particles, etc.\
 could possibly cross $B_1$ (Lieb-Robinson time).
 
 Before we embark on a consideration of realistic time and length scales, let
 us settle the required ratio of sizes of $B_1$ and $B_2$. If $B_1$, $B_2$ were
 non-interacting, a mixed microcanonical state from an energy shell of the full
 bath would result in a canonical state on $B_1$ only if the density of states of 
 $B_2$ was well described by an exponential $e^{\beta E}$ over the full
 energetic spectrum of $B_1$ \cite{goldstein06}. This, however, requires that
 ($B_1$ and $B_2$ are indeed structurally identical systems) $B_2$ must be larger
 than  $B_1$ by at least a factor of ten or so. Thus, even if the ETH nicely
 applies and the actual interaction between $B_1$, $B_2$ is negligible (for
 details cf.\ \cite{riera12}), the ratio must be at least, say, $1$:$10$ to get
 a reduced canonical state on $B_1$.
 
To grasp the gist of the problem of relevant scales, consider standard experiments
in the context of the Jarzynski relation (which is a prime example of a fluctuation
relation): the unfolding of proteins (cf., e.g., \cite{liphardt02}). These
experiments are done in aqueous solution. While the (conceptual) Lieb-Robinson
speed of water is hard to determine, it must at least exceed the respective
speed of sound, ca.\ $1400 \, \text{m/s}$. The time scales on which these
experiments are performed are on the order of seconds. Thus, for the argument of
Iyoda {\it et al.}\ to hold, the near bath $B_1$ would have to be ca.\ $1.4 \,
\text{km}$ large. Consequently, the far bath $B_2$ would have to be $14 \,
\text{km}$ large. The vessels in which these experiments are actually performed
are, however, on the order of centimeters. Thus, this scheme cannot explain the
applicability of a standard fluctuation theorem in a standard setting. Similarly,
to explain the validity of the second law according to \cite{iyoda17} for a cup of
coffee that cools down during, say, ten minutes, one would need ca.\ $200 \,
\text{km}$ of undisturbed air around the cup as $B_1$, which amounts to a volume
of air on the scale of $2000 \, \text{km}$ as $B_2$.
These facts are in sharp contrast to statements like (literal
quotation): ``Our result reveals a universal scenario that the second law emerges
from quantum mechanics, \ldots'' and ``\ldots, in this Letter we rigourously derive
the second law of thermodynamics for isolated quantum system in pure states.'' Note
that \cite{iyoda17} does not offer any discussion of an upper bound to the
Lieb-Robinson based on realistic sizes of concrete systems, but exclusively
focuses on the hypothetical limit of infinitely large baths.
 
While the numerics in \cite{iyoda17} appear to confirm the impact of the
theoretical reasoning on concrete quantum dynamics at first sight, more thorough
analysis reveals that this is not the case. To elucidate this, we re-did the
numerics in \cite{iyoda17} but (i) applied  the original parameters to {\it both}
dynamical quantities considered in \cite{iyoda17}, (ii) checked the effect of
changing one parameter, and (iii) additionally computed a simple observable that
indicates how much the system is evolving at all.
 
The first claim in \cite{iyoda17} specifically concerns the positivity of an
``average entropy production'', $\langle \sigma \rangle$. Indeed, for the example
parameters chosen in \cite{iyoda17}, this entropy production turns out to be
positive at all times, cf.\ Fig 1. Consider, however, the following variation:
Change all on-site potentials from $\omega = 1$ to $\omega = -50$. Since the model
conserves particle numbers and the initial state has precisely five bosons, this
only amounts to an energy shift of the full spectrum,
i.e., adding 
original Hamiltonian.
Hence, this variation can neither change the Lieb-Robinson time nor the applicability
of any physically reasonable form
of the ETH. (Moreover, it does not even change the effective inverse temperature
$\beta$.) It does, however, change the local energy absorbed (or released) by the
bath, thus it changes the entropy production $\langle \sigma \rangle$. The numerical
result for this situation is displayed in Fig.\ 1, together with the results for the
original on-site potentials and the occupation probability $\langle n_0(t) \rangle$
of the system site. The entropy production is clearly negative as soon as the
system leaves its initial state at all, i.e., as soon as  $\langle n_0(t) \rangle$
deviates significantly from its initial value $\langle n_0(0) \rangle = 1$.
This occurs irrespective of the tentative Lieb-Robinson time $\tau \approx 1$. Note that $\omega = -50$ is not a unique choice. For example, a
larger positive $\omega$ also yields a negative $\langle \sigma \rangle$ if the initial 
state has the system site empty.

\begin{figure}
\includegraphics[width=0.8\columnwidth]{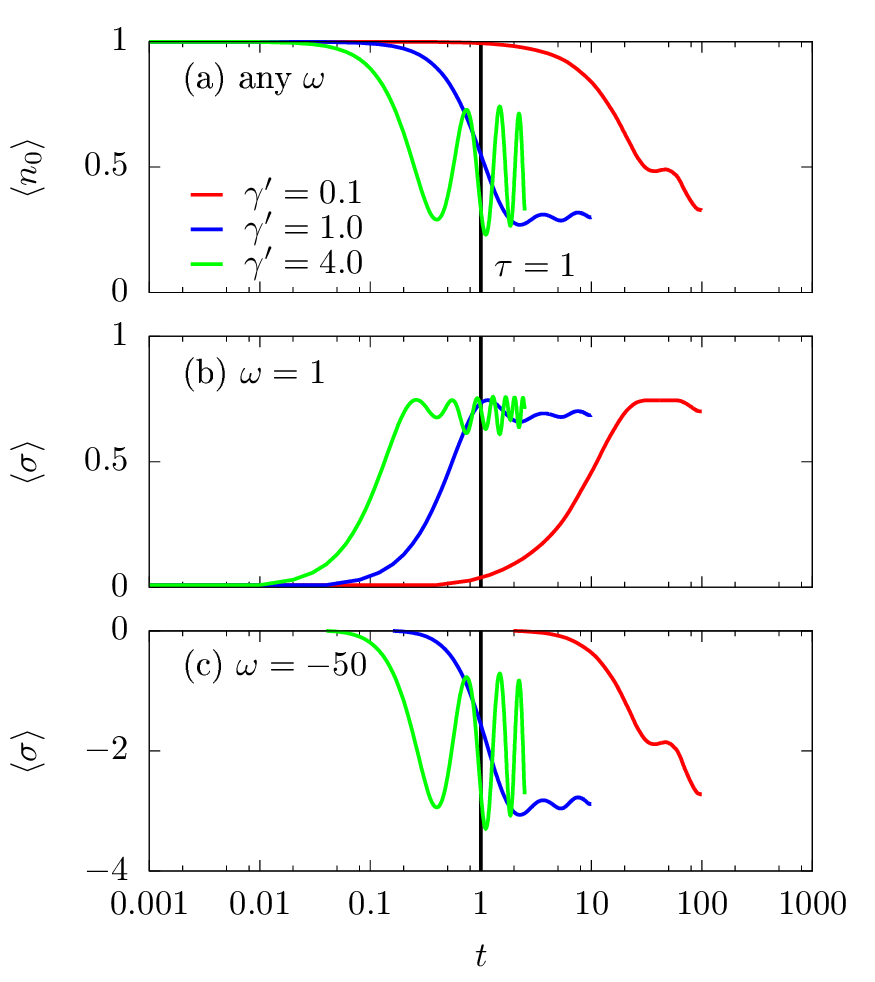}
\caption{(Color online) Time dependence of the (a) occupation probability
$\langle n_0 \rangle$ of the system site (any $\omega$), (b) average entropy
production $\langle \sigma \rangle$ for on-site potential $\omega = 1$
(original model), and (c) average entropy production  $\langle \sigma \rangle$
for on-site potential $\omega = -50$, at various interaction strengths
$\gamma'$.}
\label{fig1}
\end{figure}
\begin{figure}
\includegraphics[width=0.8\columnwidth]{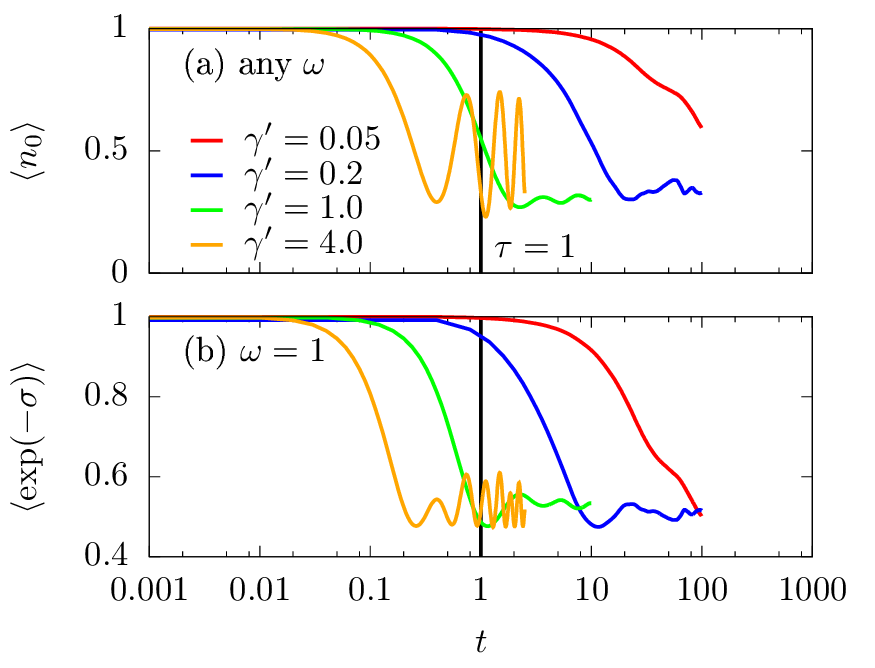}
\caption{(Color online) Time dependence of the (a) occupation probability
$\langle n_0 \rangle$ of the system site (any $\omega$) and (b) ``integral
fluctuation quantity'' $ \langle \exp(-\sigma) \rangle$ for on-site
potential $\omega = 1$. }
\label{fig2}
\end{figure} 

The second claim addresses an integral fluctuation theorem which states $\langle
e^{-\sigma} \rangle=1 $, where $\sigma$ depends on the bath Hamiltonian and
the full system state. Of course, at $t=0$ the theorem holds by construction.
In Fig.\ 2 we (re-)plot $\langle e^{-\sigma} \rangle$ for the interaction
strengths, $\gamma'$, as addressed in \cite{iyoda17} in the context of the integral
fluctuation theorem, but add data also for the interaction strength $\gamma'=4$,
the latter is addressed in \cite{iyoda17} only in the context of the afore
mentioned entropy production. Again, we also plot the respective
$\langle n_0(t) \rangle$. Obviously, $\langle e^{-\sigma} \rangle \approx 1$ only
holds to the extend to which the system remains in its initial state. Hence, the
validity of the integral fluctuation theorem at short times is rather trivial. It
breaks down as soon as the system ``does something'', irrespective of the
$\gamma'$-independent Lieb-Robinson time  $\tau \approx 1$. According
to a further claim in \cite{iyoda17}, the mere scaling of $\langle e^{-\sigma} \rangle$
as $1-ct^2$ below the Lieb-Robinson time supposedly implies the applicability 
of the respective theorem. However, while a scaling of any quantity $a(t)$ as $a(0)
+ c t^2$ at very short times is in general all but surprising in quantum mechanics
(e.g.\ \cite{chiu77}), $\langle e^{-\sigma} \rangle$ even fails to show this scaling
for $\gamma'=4$ up to the Lieb-Robinson time $\tau \approx 1$.

\newpage



{\bf Reply to arXiv:1712.05172 }

In the following we reply to the reply by E. Iyoda, K. Kaneko, and T. Sagawa 
(arXiv:1712.05172) to our previous comment on Phys. Rev. Lett. 119,
100601 (2017). 




\textcolor{OliveGreen}{Iyoda {\it et al.} replied \cite{Iyoda2017a} to a comment 
by Gemmer {\it et al.} \cite{Gemmer2017} on the arXiv. The paper at hand is our 
response to this reply \cite{Iyoda2017a}. For completeness, we reprint below 
the text of \cite{Iyoda2017a} in black lettering. We insert our responses to 
specific sections in green lettering. These responses conclude our discussion 
on the arXiv.}


Gemmer {\it et al.} in their Comment~\cite{Gemmer2017} make criticisms on our 
Letter~\cite{Iyoda2017}, while they agree that our result is mathematically 
sound. Their arguments are summarized as (i) the Lieb-Robinson (LR) time 
$\tau_{\mathrm{LR}}$ is too short and ``unphysical'' for some examples, (ii) the 
average entropy production $\langle\sigma\rangle$ becomes negative for some 
parameters, and (iii)  in our numerical simulation, the integral fluctuation 
theorem (IFT) $\langle e^{-\sigma}\rangle=1$ holds only while the system remains 
in its initial state. Here we discuss that (i) and (ii) are not justified, but 
that (iii) is indeed a subtle point because of the large finite-size effect.


\textcolor{OliveGreen}{{\it Comment to the previous section:} The criticism in 
\cite{Gemmer2017} does not consist of three different points. It is essentially 
just one point, namely, that the range of the validity of the theorems in 
\cite{Iyoda2017} is physically extremely limited, rather than just not 
comprising ``some examples''. This limitation, which arises from finite bath 
sizes, is neither discussed nor clearly stated in \cite{Iyoda2017}. Instead,
a numerical example is presented to supposedly illustrate the validity of the 
theorems for relevant time scales. What is called points (ii) and (iii) above, 
are really only numerical observations, which elucidate that the theorems in 
\cite{Iyoda2017} are only trivially valid as long as the system does nothing.}


\underline{Reply to (i).} 
We agree that the LR times of  their examples (proteins in water and coffee in 
air) are very short compared to their time scales (Brownian motion and daily 
life). However, these situations are clearly not relevant to our theory in 
\cite{Iyoda2017}, which is for isolated quantum systems where the bath is 
initially in a pure state (specifically in an energy eigenstate). In fact, the 
setups discussed in \cite{Gemmer2017} are far from isolated and pure. In other 
words, in their setups the fluctuation theorem is not emergent from quantum 
mechanics, but is simply a consequence of the conventional scenario based on 
classical stochastic dynamics. Our result in \cite{Iyoda2017} implies that if 
air or water was initially in an energy eigenstate, then the IFT would hold only 
within such a very short time scale.


\textcolor{OliveGreen}{{\it Intermediate comment:} This is not correct: an IFT may  
still hold much longer due to alternative mechanisms, e.g., as suggested in 
\cite{schmidtke2017}.}


We emphasize that our main focus is on ideally-isolated artificial quantum 
systems such as ultracold atoms, as explicitly stated in  \cite{Iyoda2017}. In 
fact, the LR time is  reasonably long for ultracold atoms, compared to the time 
scale of real experiments. For example, in a typical experiment 
\cite{Cheneau2012}, the experimental time scale is $\hbar/J$ and the LR time is  
$\tau_\mathrm{LR}\sim l\hbar/J$, where $J$ is the tunneling amplitude and $l$ 
is the side length of ${\rm B}_1$ in~\cite{Iyoda2017}. 


\textcolor{OliveGreen}{{\it Comment to the previous section:} The conceptual framework 
of the paper is that of quantum typicality, the eigenstate thermalization 
hypothesis, and Lieb-Robinson bounds. The first two concepts are notable and 
relevant precisely because they potentially explain fundamental features of 
thermodynamical behavior (like equilibration and thermalization) 
for the full range of systems, i.e, from microscopic ultracold atom experiments 
all the way to macroscopic daily life physics such as cups of coffee. As cited 
in \cite{Gemmer2017}, a number of statements in \cite{Iyoda2017} imply the same 
validity range and  impact for the findings in \cite{Iyoda2017}. Here we quote 
yet another on. In the introduction of  \cite{Iyoda2017} it reads: ``In this 
sense, a fundamental gap between the microscopic and the macroscopic worlds has 
not yet been bridged: How does the second law emerge from pure quantum states?''
 We leave it to the reader to decide whether the fact that the results of 
\cite{Iyoda2017} are truly conceptually limited to specific possible future 
ultracold atom experiments is made sufficiently transparent in \cite{Iyoda2017}.
}


\underline{Reply to (ii).} 
The reason why the authors of~\cite{Gemmer2017} observed the negative entropy 
production for some parameters is that their choice of the initial eigenstate 
is not thermal for such cases, as detailed below. We note that our theory 
\cite{Iyoda2017} states that the second law and the IFT hold if the initial 
eigenstate is thermal.

In the hard-core boson model in~\cite{Gemmer2017,Iyoda2017}, the total Hilbert 
space of bath B is divided into particle-number sectors, labeled by $N$. An 
energy eigenstate is thermal only if  it is in a sector whose $N$ is close to 
the average particle number $N^\ast$ in the canonical ensemble, because the 
strong eigenstate thermalization hypothesis (ETH) is valid only within each 
particle number sector~\cite{DAlessio2016}. We note that the weak ETH is true 
without dividing the energy shell to particle-number sectors. Because $N^\ast 
\simeq 15.9$ with $\omega = -50$ and $\beta = 0.1$, the choice of $N=4$ in 
\cite{Gemmer2017}  is far from $N^\ast$, which makes the initial eigenstate 
athermal.

\begin{figure}[t]
\begin{center}
\includegraphics[width=\linewidth]{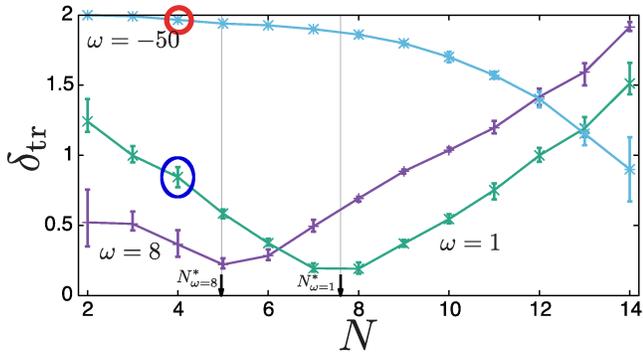}
\end{center}
\label{Main_fig1}
\caption{
The $N$-dependence of $\delta_\mathrm{tr}$. The Hamiltonian and the lattice are 
the same as in \cite{Gemmer2017, Iyoda2017}. The parameters are given by 
$\omega=1,8,-50$, $g=0.1$, and $\beta=0.1$. $N^\ast_{\omega=1}$ and 
$N^\ast_{\omega=8}$ are the average particle numbers in the canonical ensemble 
for $\omega=1,8$, respectively. For each data point, 10 energy eigenstates are 
sampled, and the error bar represents their standard deviation. The red circle 
indicates the parameters used in \cite{Gemmer2017} ($\omega=-50$ and $N=4$).
\textcolor{OliveGreen}{The blue circle indicates the parameters used in 
\cite{Iyoda2017} ($\omega=1$ and $N=4$).}
}
\end{figure}

To directly show this, we calculated the trace norm between the reduced density 
operators of an energy eigenstate $\ket{E_i}$ and the corresponding canonical 
ensemble: $\delta_\mathrm{tr}:=\|\mathrm{tr}_{\mathrm{B}_2}[\ket{E_i}\bra{E_i}]
-\mathrm{tr }_{\mathrm{B}_2}[\hat{\rho}_{\mathrm{B},\mathrm{can}}]\|_1$. Figure 
1 shows the $N$-dependence of $\delta_\mathrm{tr}$, where we take 
$\mathrm{B}_1$ as the $2\times 2$ lower-left sites of bath B. As shown in 
Fig.~1, $\delta_\mathrm{tr}$ takes a smaller value when $N$ is closer to $N^*$. 
For   $\omega=-50$ and $N=4$ (red circle), $\delta_\mathrm{tr}$ is 
large and $\ket{E_i}$ is not at all thermal. 

We have also confirmed that the entropy production is positive for a broad range 
of parameters, as long as the initial eigenstate is thermal. We calculated 
$\langle \sigma \rangle$ at $t=\tau_{\rm LR}$ for $\omega = \pm 1, \pm 2, \pm 
4, \pm 8, \pm 16, \pm 32, - 50$, $\beta =0.1,0.3$, $\gamma' = 
0.05,0.1,0.4,1.0,4.0$, $g=0.1,0.4$, and $N=4$. We found that $0.0076 \leq 
\langle \sigma \rangle \leq 1.84$ if $\delta_\mathrm{tr} < 0.3$ (thermal),
while $-8.86 \leq \langle \sigma \rangle \leq 0.095$ if $\delta_\mathrm{tr} > 
1.7$ (athermal).


\textcolor{OliveGreen}{{\it Comment to the previous section:} While a numerical  
agreement with the IFT would be remarkable because it is requires $ \langle 
\exp(-\sigma) \rangle$  to equal precisely unity, an agreement with the 
``entropy production theorem'' is much less remarkable, even for a ``broad range 
of parameters'', since it only requires $\langle \sigma \rangle$ to be 
non-negative. The latter may simply occur by chance, cf.\ below. Furthermore, 
as the title indicates, \cite{Iyoda2017} is primarily about the fluctuation 
theorem, rather than about the much weaker positivity of entropy production.
\\
First of all, the reply at hand by Iyoda {\it et al.} indicates that 
``thermality'' of the initial state acutely requires an involved fine tuning of 
model parameters that may possibly work out at some specific instance. This 
appears to be true even in the limit of infinitely large baths. Again, we leave 
it to the reader to decide whether this fact is clearly communicated in 
\cite{Iyoda2017}.
\\
Moreover, in the original paper \cite{Iyoda2017} the error bound for the entropy 
production theorem is given by $\epsilon_{2nd}$ [Eq.(3)]. In the reply at hand
Iyoda {\it et al.} surprisingly  argue based on the smallness of a parameter 
$\delta_\mathrm{tr}$ which does not even appear in the original paper. 
Apparently, $\delta_\mathrm{tr}$ has to be small to guarantee ``thermality'' of 
the initial state. It is, however, entirely unclear how small  
$\delta_\mathrm{tr}$ has to be to indicate sufficient thermality. In the reply 
at hand Iyoda {\it et al.} classify (without further justification)  
$\delta_\mathrm{tr} < 0.3$ (thermal) and $\delta_\mathrm{tr} > 1.7$ (athermal).
While the example in \cite{Gemmer2017} ($\omega =-50, N=4$) yields 
$\delta_\mathrm{tr} \approx 2$ and thus barely results as ``athermal'', the 
original example in \cite{Iyoda2017} ($\omega=1, N=4$) yields 
$\delta_\mathrm{tr} \approx 0.7$ (blue circle \cite{blue}), i.e., does not 
qualify as ``thermal'' either, according to the above {\it ad hoc} 
classification. Given that $\delta_\mathrm{tr}$  is meant to quantify the 
distance of the local initial bath state from a true thermal state, it should, 
to be a bit more quantitative, at least be compared to the corresponding norm  
of the thermal state itself. The latter equals unity (i.e., the 1-Schatten-norm 
of a density matrix). So for ``closeness''  to the thermal state one should 
require $\delta_\mathrm{tr} \ll 1$. This, however, is hardly reached anywhere 
in all the many parameter combinations displayed in Fig.\ 1,  and surely not 
for the original example (blue circle). This suggests that, due to the bath 
being far too small, the presented model is for no parameter setting really 
within the regime of validity of the proclaimed theorem and the positivity of 
the entropy production is a coincidence which is not truly enforced by the 
validity of the theorem.}


\underline{Reply to (iii).}  
In~\cite{Iyoda2017}, we concluded that the IFT nontrivially holds in the short 
time regime based on the fact that the quantitative error evaluation, scaling 
of $\propto t^2$, is consistent with the prediction by the LR argument (the 
inset of Fig.~3 in \cite{Iyoda2017}). On the other hand, the authors of 
\cite{Gemmer2017} argued that this scaling is just a general property of quantum 
systems.  
 
After careful consideration with some additional numerical simulation, we have 
to admit that it is hard to conclude whether the $t^2$-scaling comes from the 
LR argument or not based on our numerical data, because the finite-size effect 
is very large within the numerically-accessible system size. We expect, 
however, that the IFT can be verified more clearly by using ultracold atoms, 
with which the system size can be much bigger than numerics. We note that the 
time range where the IFT holds will linearly increase with $l$. If one can take 
$l \sim 10^2$, the LR time becomes a hundred times of the experimental time 
scale, which can be realized with the current or the near-future technologies.

In addition, we remark that $|\gamma^\prime|$ should not be taken too large to 
verify the IFT, because the larger $|\gamma^\prime|$ is, the more the condition 
(8) in \cite{Iyoda2017} is violated. In this respect, the emphasis in 
\cite{Gemmer2017} on the failure of the IFT for $\gamma^\prime = 4$ is 
misleading.


\textcolor{OliveGreen}{{\it Comment to the previous section:} The justification of the 
fluctuation theorem is the major content of \cite{Iyoda2017} as already the 
title indicates. Iyoda {\it et al.} present a numerical example (instead of a 
discussion of actual finite bath sizes) to support the applicability of the 
theorem. In \cite{Gemmer2017} we demonstrated that the theorem only holds if 
the system does nothing. We further argued that a quadratic time dependence of 
$ \langle \exp(-\sigma) \rangle$ does not indicate the validity of the theorem. 
In the reply at hand Iyoda {\it et al.} admit that a quadratic time dependence 
of $\langle \exp(-\sigma) \rangle$ does not indicate the validity of the 
theorem. However, they do not comment at all on the fact that the theorem only 
holds as long as the system does nothing. But this is the most important point. 
As long as this point stands, the numerical example does not support the 
applicability and relevance of the theorem in any way. (As opposed to the statement 
in the abstract of \cite{Iyoda2017} (literal quotation): ``We confirmed our theory 
by numerical simulation of hard-core bosons, and observed dynamical crossover
from thermal fluctuations to bare quantum fluctuations''.) Since, to repeat, the 
regime of applicability of the theorem is extremely narrow, this finding is 
hardly surprising. It only concretely reflects the general fact that the range 
of validity of the proclaimed theorem is very small compared to the factually 
immensely large applicability of the second law or fluctuation theorems. This 
is 
true irrespective of any hypothetical models or possible future experiments on 
cold atoms.}


\begin{acknowledgments}
\textcolor{OliveGreen}{We acknowledge fruitful discussions on this subject with D. 
Schmidtke. This work has been funded by the Deutsche Forschungsgemeinschaft 
(DFG) - GE 1657/3-1; STE 2243/3-1.}
\end{acknowledgments}


\newpage

\end{document}